\documentclass[aps,pccp,preprint,superscriptaddress]{revtex4-1} 
\usepackage{graphicx} 
\usepackage{amsmath} 
\usepackage[version=3]{mhchem} 
\usepackage{longtable} 
\usepackage{filecontents} 

\begin{document} 

\title{Assessment of asymptotically corrected model potential scheme for charge-transfer-like excitations in oligoacenes} 

\author{Wei-Tao Peng} 
\affiliation{Department of Physics, National Taiwan University, Taipei 10617, Taiwan} 

\author{Jeng-Da Chai} 
\email[Author to whom correspondence should be addressed. Electronic mail: ]{jdchai@phys.ntu.edu.tw} 
\affiliation{Department of Physics, National Taiwan University, Taipei 10617, Taiwan} 
\affiliation{Center for Theoretical Sciences and Center for Quantum Science and Engineering, National Taiwan University, Taipei 10617, Taiwan} 
\affiliation{Physics Division, National Center for Theoretical Sciences (North), National Taiwan University, Taipei 10617, Taiwan}

\date{\today} 

\begin{abstract} 

We examine the performance of the asymptotically corrected model potential scheme on the two lowest singlet excitation energies of acenes with different number of linearly fused benzene rings (up to 5), employing both the 
real-time time-dependent density functional theory and the frequency-domain formulation of linear-response time-dependent density functional theory. The results are compared with the experimental data and those calculated 
by long-range corrected hybrid functionals and others. The long-range corrected hybrid scheme is shown to outperform the asymptotically corrected model potential scheme for charge-transfer-like excitations. 

\end{abstract} 

\maketitle

\section{Introduction} 

Over the past two decades, time-dependent density functional theory (TDDFT) \cite{RG} has been a popular method for the study of excited-state and time-dependent properties of large systems, due to its favorable balance 
between accuracy and efficiency \cite{TDDFT,TDDFT2}. However, the exact exchange-correlation (XC) potential $v_{xc}({\bf r}, t)$ in TDDFT remains unknown, and needs to be approximated for practical applications. 

For a system subject to a slowly varying external potential, the most popular approximation for $v_{xc}({\bf r}, t)$ is the adiabatic approximation, 
\begin{equation} 
v_{xc}({\bf r}, t) \approx \frac{\delta E_{xc}[\rho]}{\delta \rho({\bf r})} \bigg |_{\rho({\bf r})=\rho({\bf r},t)}, 
\label{eq:vxc} 
\end{equation} 
where $v_{xc}({\bf r}, t)$ is approximated by the functional derivative of the XC energy functional $E_{xc}[\rho]$ evaluated at the instantaneous density ${\rho({\bf r},t)}$. In the adiabatic approximation, memory effects, 
whereby $v_{xc}({\bf r}, t)$ may depend on the density at all previous times ($t' < t$), are completely neglected. Surprisingly, results obtained from the adiabatic approximation can be accurate in many cases, even if the 
system considered is not in this slowly varying regime. However, as the exact $E_{xc}[\rho]$, which appears in both Kohn-Sham density functional theory (KS-DFT) \cite{KS} (for ground-state properties) and adiabatic 
TDDFT (for excited-state and time-dependent properties), has not been known, the development of a generally accurate density functional approximation for $E_{xc}[\rho]$ remains an important and challenging 
task \cite{DFTReview2,DFTreview}. 

Functionals based on the localized model XC holes, such as the local density approximation (LDA) and generalized gradient approximations (GGAs), are reliably accurate for applications governed by short-range XC effects, 
such as low-lying valence excitation energies. However, they can produce erroneous results in situations where the accurate treatment of nonlocal XC effects is important. In particular, some of these situations occur in the 
asymptotic regions of atoms and molecules, where the LDA or GGA XC potential exhibits an exponential decay, instead of the correct $-1/r$ decay. Accordingly, LDA and GGAs (i.e., semilocal density functionals) severely 
underestimate high-lying Rydberg excitation energies \cite{R9,R10,R11,R12}, and completely fail for charge-transfer (CT) excitation energies \cite{R12,R15,R16,R17,R18,R19} and excitations in completely symmetrical 
systems where no net CT occurs \cite{R21}. 

Aiming to resolve the asymptote problem, long-range corrected (LC) hybrid functionals \cite{R23,R24,BNL,wB97X,wB97X-D,wM05-D,LC-D3} and asymptotically corrected (AC) model potentials \cite{LB94,LBa,AA,AC1,Tozer}, 
which are two distinct density functional methods with correct asymptotic behavior, have been actively developed over the past few years. The LC hybrid scheme, which adopts 100\% Hartree-Fock (HF) exchange for 
long-range electron-electron interactions, thereby provides an AC XC potential (i.e., a local multiplicative XC potential), when the optimized effective potential (OEP) method is employed \cite{OEP1,OEP2,OEP3,DFTReview2}. 
Similar to other orbital-dependent XC energy functionals, a generalized Kohn-Sham (GKS) method (i.e., using orbital-specific XC potentials) has been frequently employed in the LC hybrid scheme to circumvent the 
computational complexity of the OEP method, as the density, energy, and highest-occupied orbital energy obtained from the GKS method are generally similar to those obtained from the OEP method \cite{OEP3,DFTReview2}. 
In our recent work, the performance of the LC hybrid scheme (i.e., using the GKS method) and AC model potential scheme has been examined on a very wide range of applications \cite{LCAC}. In particular, we have shown 
that LC hybrid functionals could be reliably accurate for various types of excitation energies, including valence, Rydberg, and CT excitation energies, in the frequency-domain formulation of linear-response TDDFT 
(LR-TDDFT) \cite{R4}. Nevertheless, due to the inclusion of long-range HF exchange, the LC hybrid scheme can be computationally expensive for large systems. 

On the other hand, in the AC model potential scheme, an AC XC potential is directly modeled, maintaining computational complexity similar to the efficient semilocal density functional methods. However, as most popular 
AC model potentials are found {\it not} to be functional derivatives \cite{Staroverov,Staroverov2b}, the associated XC energies and XC kernels (i.e., the second functional derivative of $E_{xc}[\rho]$) are not well-defined. 
Accordingly, an adiabatic LDA or GGA XC kernel (i.e., {\it not} a self-consistent adiabatic XC kernel) has been frequently adopted for the LR-TDDFT calculations using the AC model potential scheme. Such combined 
approaches have been shown to perform well for both valence and Rydberg excitations, but very poorly for CT excitations \cite{LCAC,R20,R21,R22new,new1}, due to the lack of a space- and frequency-dependent 
discontinuity in the adiabatic LDA or GGA XC kernel adopted in LR-TDDFT \cite{R22}. Nevertheless, it remains unclear whether the AC model potential scheme can accurately describe CT or CT-like excitations, when 
a self-consistent adiabatic XC kernel (if available) is adopted in LR-TDDFT. 

To circumvent this problem, in this work, we examine the performance of the AC model potential scheme on various types of excitation energies in the real-time formulation of TDDFT (RT-TDDFT), since LR-TDDFT is typically 
a good approximation to RT-TDDFT. As an absorption spectrum (and hence, excitation energies) can be obtained by explicitly propagating the time-dependent Kohn-Sham (TDKS) equations, the knowledge of the XC kernel 
is not needed within the framework of RT-TDDFT. Particularly, we like to address if the AC model potential scheme is able to accurately describe CT-like excitations in RT-TDDFT, which to the best of our knowledge has never 
been addressed in the literature. The rest of this paper is organized as follows. In section II, we describe our test sets and computational details. The excitation energies calculated by the AC model potential scheme, the LC 
hybrid scheme, and others in both RT-TDDFT and LR-TDDFT are compared with the experimental data and the results obtained from a highly accurate {\it ab initio} method in section III. Our conclusions are given in section IV.

\section{Test Sets and Computational Details} 

\begin{figure}
\includegraphics[scale=0.8]{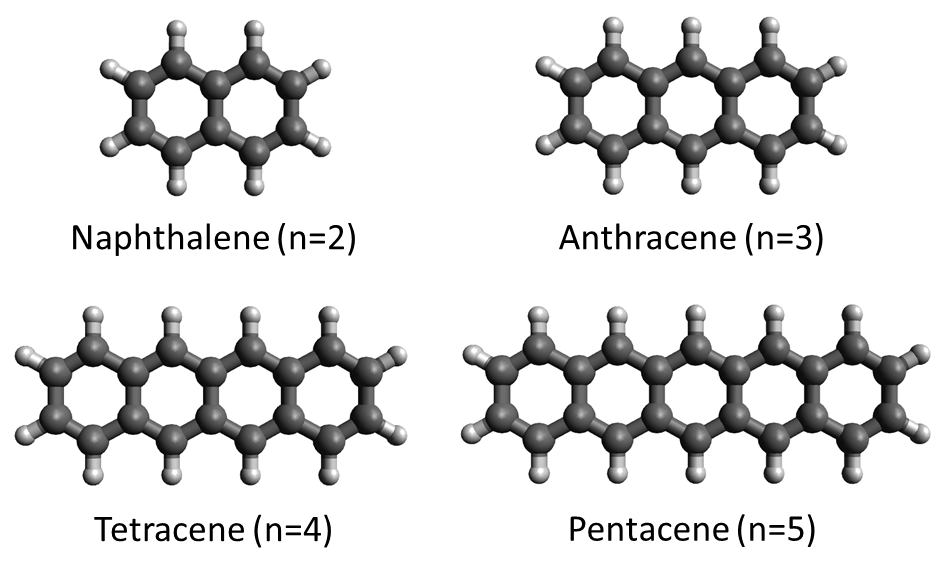} 
\caption{\label{fig:1} Structures of the $n$-acenes investigated.} 
\end{figure}

Linear $n$-acenes (C$_{4n+2}$H$_{2n+4}$), consisting of $n$ linearly fused benzene rings (see Fig.\ \ref{fig:1}), are important molecules for a variety of devices, such as organic light-emitting diodes \cite{R29}, 
solar cells \cite{R27}, and field-effect transistors \cite{R26}. Recently, the two lowest singlet $\pi \rightarrow \pi^*$ transitions of $n$-acenes, commonly labelled as the $^1L_a$ (the lowest excited state of B$_{2u}$ symmetry) 
and $^1L_b$ (the lowest excited state of B$_{3u}$ symmetry) states in Platt's nomenclature \cite{Lab}, have received considerable attention \cite{R7,R30,R31,R32,R33,R34,new2,new3,new4}. 

The $^1L_a$ state is dominated by a transition between the highest occupied molecular orbital (HOMO) and lowest unoccupied molecular orbital (LUMO), i.e., the HOMO $\rightarrow$ LUMO transition, with polarization 
along the molecular short axis, while the $^1L_b$ state is dominated by a combination of the two nearly degenerate configurations HOMO $-$ 1 $\rightarrow$ LUMO and HOMO $\rightarrow$ LUMO $+$ 1, with polarization 
along the molecular long axis. From a valence-bond point of view, the $^1L_a$ state is mainly ionic in character, whereas the $^1L_b$ state is mainly covalent in character \cite{R30,R31}. 

Kuritz {\it et al.} described the $^1L_a$ state as a CT-like excitation \cite{R34}, and showed that through a unitary transformation, the coupling between the HOMO and LUMO is weak, supporting the surmise of Richard and 
Herbert that the $^1L_a$ state has CT character in disguise \cite{R33}. Note that as the $^1L_a$ state is not a pure CT excitation (i.e., for well-separated donor-acceptor systems) \cite{R15,R16}, the terminology ``CT-like'' 
may be controversial \cite{new2,new3,new4}. However, in this work, the $^1L_a$ state is regarded as a CT-like excitation, as suggested in Refs.\ \cite{R33,R34}. 
By contrast, the $^1L_b$ state is a valence excitation with substantial double-excitation character \cite{R33}. 
In LR-TDDFT, although LDA and GGAs can accurately predict the $^1L_b$ excitation energy, they substantially underestimate the $^1L_a$ excitation energy \cite{R7,R30,R31,R32,R33,R34,new2}. 
On the other hand, LC hybrid functionals are reliably accurate for the $^1L_a$ state, but less accurate for the $^1L_b$ state \cite{R7,R32,R33,R34,new2}. Note that the efficient AC model potential scheme 
has never been examined on the $^1L_a$ and $^1L_b$ states of $n$-acenes in the literature. 

To examine the performance of several density functional methods on various types of excitation energies in both RT-TDDFT and LR-TDDFT, the $^1L_a$ and $^1L_b$ states of $n$-acenes (up to 5-acene) are adopted as 
our test sets. For the LR-TDDFT calculations, we adopt LDA \cite{LDAX,LDAC}, PBE \cite{PBE} (a popular GGA functional), and LB94 \cite{LB94} (a popular AC model potential) with the 6-31+G(d,p) basis set, 
to calculate the $^1L_a$ and $^1L_b$ excitation energies on the ground-state geometries of $n$-acenes obtained at the $\omega$B97X/6-31G(d) level \cite{wB97X} with a development version of \textsf{Q-Chem 4.0} \cite{R36}. 
For the LDA and PBE calculations, the adiabatic LDA and PBE XC kernels (i.e., the second functional derivatives of the LDA and PBE XC energy functionals, respectively) are adopted, respectively. For the LB94 calculations, 
the adiabatic LDA XC kernel is adopted, due to the lack of a self-consistent adiabatic XC kernel for the LB94 model potential. 

The RT-TDDFT calculations, employing the adiabatic LDA, PBE, and LB94 XC potentials, are performed with the program package \textsf{Octopus 4.0.1} \cite{R35}. 
The system, which is described on a real-space grid with a 0.2 $\AA$ spacing, is constructed by adding spheres created around each atom, of radius 6 $\AA$. 
Troullier-Martins pseudopotentials are adopted to describe the complicated effects of the motion of core electrons \cite{TM}. 
To obtain spectroscopic information in RT-TDDFT, $n$-acene, which starts from the ground state, is excited via a linearly polarized delta kick in $\nu$-direction: 
\begin{equation} 
v_{ext}({\bf r}, t) = \hbar k \ \delta(t) \ {r}_{\nu}, 
\label{eq:kick} 
\end{equation} 
where ${r}_{\nu}$ is one of the Cartesian coordinates ($x$, $y$, $z$), and the perturbation strength $k$ = 0.01 bohr$^{-1}$ is adopted to obtain linear spectroscopy \cite{R35new}. 
To propagate the TDKS equations, we adopt a time step of $\Delta t$ = 0.001 $\hbar$/eV (0.658 as) and run up to 100 $\hbar$/eV (65.8 fs), which corresponds to $10^5$ time steps. 
The approximated enforced time-reversal symmetry (AETRS) algorithm is employed to numerically represent the time evolution operator \cite{aetrs}.

\section{Results and Discussion} 

\begin{figure}
\includegraphics[scale=0.6]{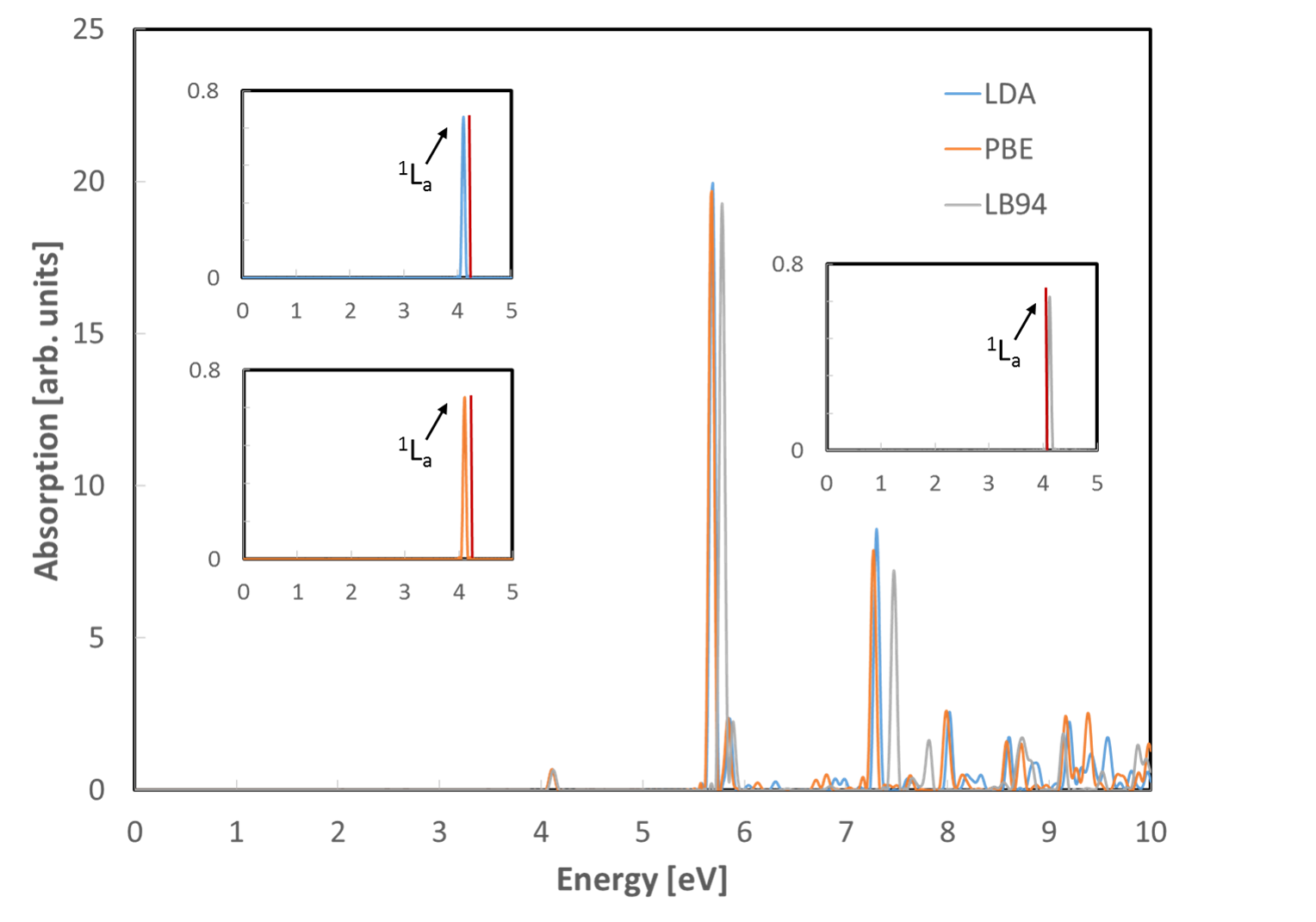} 
\caption{\label{fig:2} Absorption spectra of 2-acene calculated by various functionals in RT-TDDFT. 
Subfigures (left top: LDA; left bottom: PBE; right: LB94) show the spectra close to the position of the $^1L_a$ peaks, where the corresponding LR-TDDFT results are marked with the red lines.} 
\end{figure}

\begin{figure}
\includegraphics[scale=0.6]{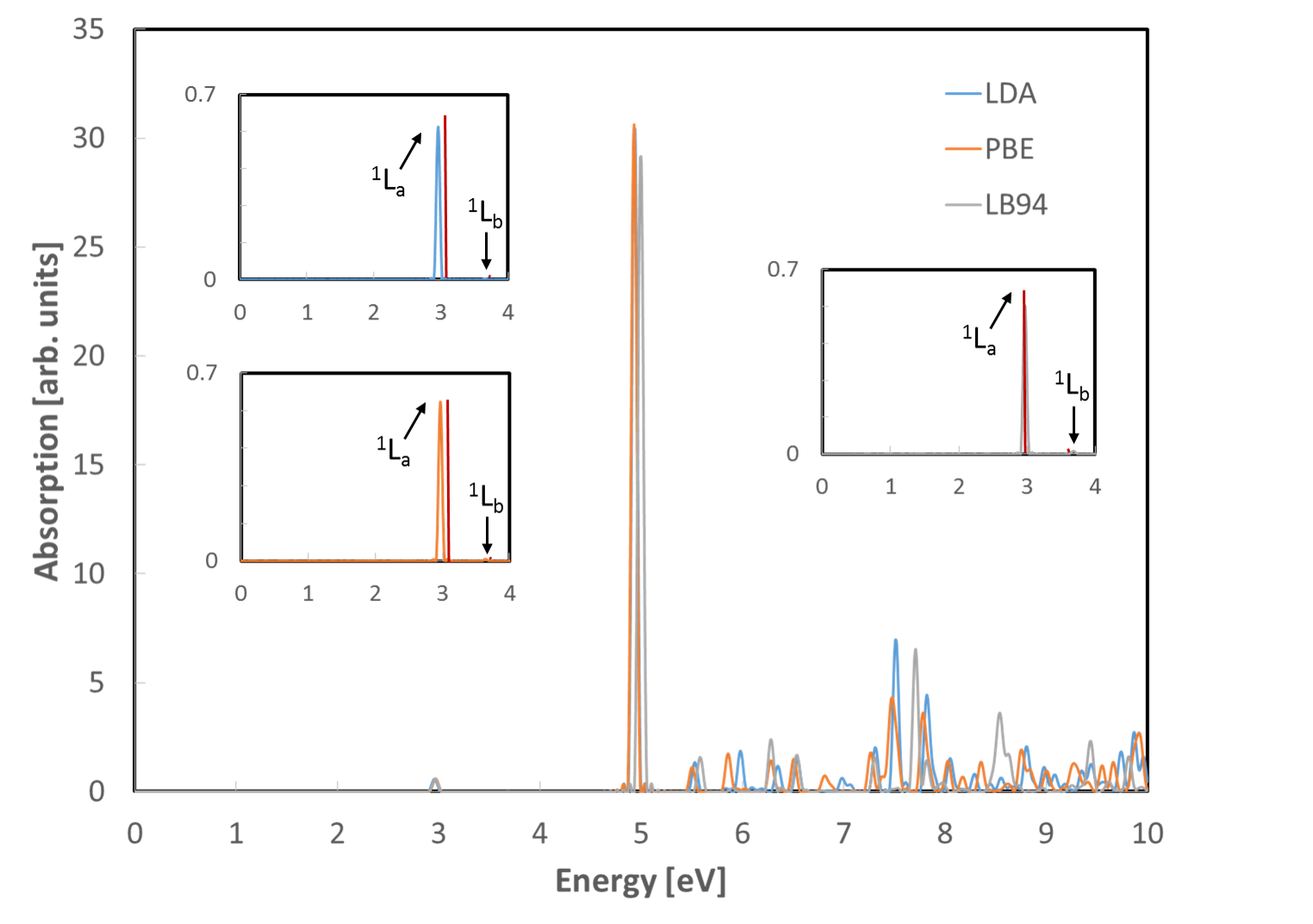} 
\caption{\label{fig:3} Absorption spectra of 3-acene calculated by various functionals in RT-TDDFT. 
Subfigures (left top: LDA; left bottom: PBE; right: LB94) show the spectra close to the position of the $^1L_a$ and $^1L_b$ peaks, where the corresponding LR-TDDFT results are marked with the red lines.} 
\end{figure}

\begin{figure}
\includegraphics[scale=0.6]{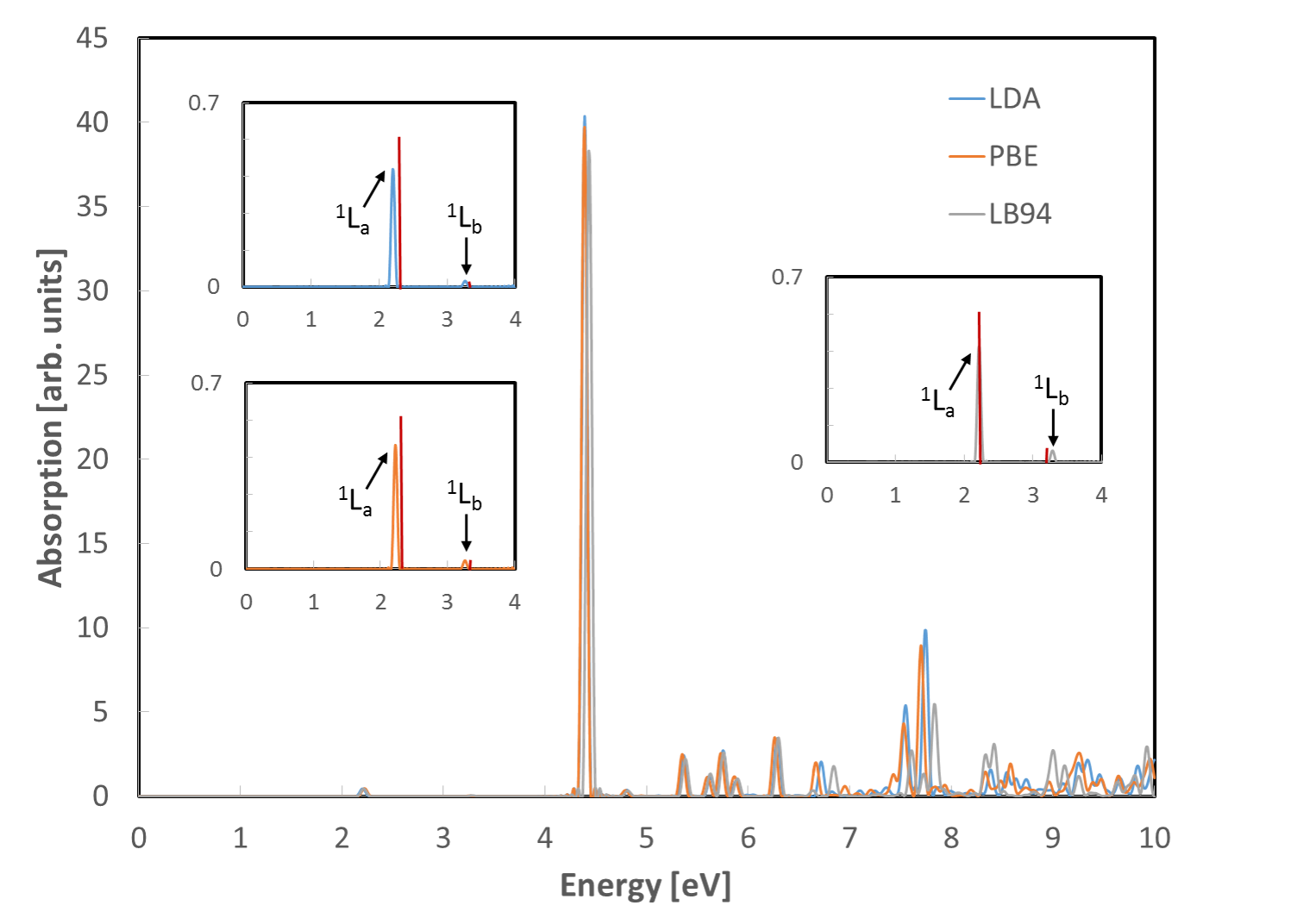} 
\caption{\label{fig:4} Same as Fig.\ \ref{fig:3}, but for 4-acene.} 
\end{figure}

\begin{figure}
\includegraphics[scale=0.6]{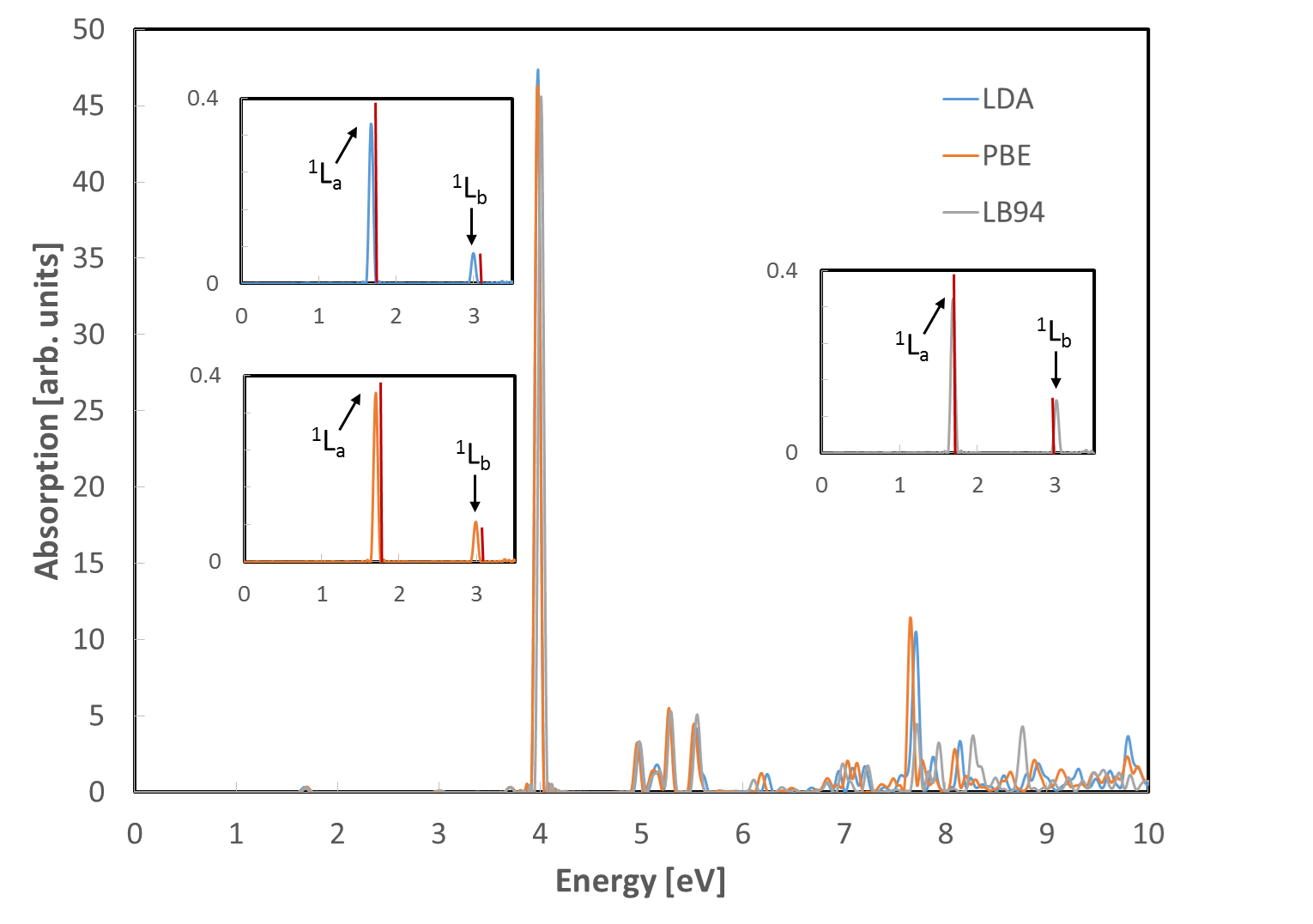} 
\caption{\label{fig:5} Same as Fig.\ \ref{fig:3}, but for 5-acene.} 
\end{figure}

The absorption spectra of $n$-acenes, calculated by LDA, PBE, and LB94 in RT-TDDFT, are plotted in Figs.\ \ref{fig:2}, \ref{fig:3}, \ref{fig:4}, and \ref{fig:5}, where the spectra close to the position of the $^1L_a$ and $^1L_b$ 
peaks are highlighted in the subfigures, and the corresponding LR-TDDFT results are marked with the red lines. Note that the $^1L_b$ state exhibits weak intensity compared with the $^1L_a$ state. For 2-acene, 
as the oscillator strengths of the $^1L_b$ state calculated by LDA, PBE, and LB94 in LR-TDDFT are found to be vanishingly small, the total propagation time adopted in our RT-TDDFT calculations may not be long enough 
to detect the $^1L_b$ state. 

For a comprehensive comparison, the $^1L_a$ and $^1L_b$ excitation energies, calculated by LDA, PBE, and LB94 in both RT-TDDFT and LR-TDDFT, are summarized in Tables {\ref{table:1}} and {\ref{table:2}}, respectively, 
where the results calculated by BNL \cite{BNL} (a popular LC hybrid functional) are taken from Ref.\ \cite{R7}, and those calculated by time-dependent coupled-cluster theory with single and double excitations (albeit with an 
approximate treatment of the doubles, CC2 \cite{CC2}) and the experimental data are taken from Ref.\ \cite{R30}. 

Owing to the CT-like character, LDA and PBE significantly underestimate the excitation energies of the ionic $^1L_a$ states in both RT-TDDFT and LR-TDDFT. Compared to the highly accurate CC2 results and experimental 
data, the errors of LDA and PBE increase with the acene length. LB94 performs similarly to LDA and PBE, indicating that the LB94 model potential is also inappropriate for CT-like excitations. 
As the LB94 results obtained from both the RT-TDDFT and LR-TDDFT calculations are very similar, the adiabatic LDA XC kernel adopted in the LR-TDDFT calculations appears to be appropriate. 
In RT-TDDFT, the failure of LB94 may be attributed to the lack of the step and peak structure in the adiabatic LB94 XC potential, which has recently been shown to be essentially important for CT excitations \cite{R38}. 
In LR-TDDFT, the failure of LB94 may be attributed to the lack of a space- and frequency-dependent discontinuity in the adiabatic LDA XC kernel adopted \cite{R19,R20,R21,R22}. 
Based on the above reasons, we expect that the CT failures may not be remedied by 
other AC model potentials exhibiting the same features as LB94 or a {\it pure} density functional whose functional derivative has the correct $-1/r$ asymptote \cite{LFA}. 
By contrast, BNL performs very well on the excitation energies of the $^1L_a$ states in both RT-TDDFT and LR-TDDFT. Fully nonlocal (i.e., orbital-dependent) functionals, in particular, LC hybrid functionals, can be essential 
for the accurate description of CT-like excitations. 

On the other hand, for the covalent $^1L_b$ states, LDA, PBE, and LB94 accurately predict the $^1L_b$ excitation energies in both RT-TDDFT and LR-TDDFT, yielding quantitative agreement with the experimental data. 
Due to the inclusion of a large fraction of HF exchange, BNL yields noticeable errors on the excitation energies of the $^1L_b$ states in both RT-TDDFT and LR-TDDFT, which may be attributed to the pronounced 
double-excitation character of the $^1L_b$ states \cite{R33}. It remains very difficult to accurately describe both the $^1L_a$ and $^1L_b$ states of $n$-acenes with existing density functionals.

\section{Conclusions} 

In this work, we have examined the performance of a variety of density functionals on the two lowest singlet $\pi \rightarrow \pi^*$ transition energies (i.e., the $^1L_a$ and $^1L_b$ states) of $n$-acenes (up to 5-acene) in 
both RT-TDDFT and LR-TDDFT. Our results have shown that the LB94 model potential performs similarly to LDA and PBE on both the $^1L_a$ (CT-like) and $^1L_b$ (valence) states. 
The excitation energies of the $^1L_a$ states are severely underestimated by LDA, PBE, and the LB94 model potential, and the errors have been shown to increase with the acene length. 
Despite its computational efficiency, our results suggest that the LB94 model potential may not accurately describe CT-like excitations in both RT-TDDFT (due to the lack of the step and peak structure in the 
adiabatic LB94 XC potential) and LR-TDDFT (due to the lack of a space- and frequency-dependent discontinuity in the adiabatic LDA XC kernel adopted). Although only the LB94 model potential has been examined 
in this work, we expect that other AC model potentials exhibiting the same features as LB94 or a pure density functional whose functional derivative has the correct asymptote may not resolve the CT problems. 

On the other hand, the LC hybrid scheme, which can be computationally expensive for large systems, is reliably accurate for the excitation energies of the $^1L_a$ states due to the inclusion of long-range HF exchange, 
but less accurate for the excitation energies of the $^1L_b$ states due to the significant double-excitation character of the $^1L_b$ states. It remains very challenging to develop a generally accurate density functional for 
the ground-state, excited-state, and time-dependent properties of large systems.

\begin{acknowledgments} 

This work was supported by the National Science Council of Taiwan (Grant No.\ NSC101-2112-M-002-017-MY3), National Taiwan University (Grant No.\ NTU-CDP-103R7855), 
the Center for Quantum Science and Engineering at NTU (Subproject Nos.:\ NTU-ERP-103R891401 and NTU-ERP-103R891403), and the National Center for Theoretical Sciences of Taiwan. 

\end{acknowledgments}

\newpage 
\begin{table*} 
\caption{\label{table:1} Excitation energies (in eV) for the $^1L_a$ states of $n$-acenes, calculated by various functionals in both RT-TDDFT and LR-TDDFT. The BNL results are taken from Ref.\ \cite{R7}, 
and the CC2 and experimental results are taken from Ref.\ \cite{R30}. The mean absolute errors (MAEs) of these methods are provided for comparisons (error $=$ theoretical value $-$ experimental value).} 
\begin{ruledtabular} 
\begin{tabular*}{\textwidth}{rrrrrrrrrrrr} 
&  \multicolumn{5}{r}{RT-TDDFT} & \multicolumn{4}{r}{LR-TDDFT} \\ 
\cline{4-7}
\cline{8-11} 
$n$-acene & Experiment & CC2 & LDA & PBE & LB94 & BNL & LDA & PBE & LB94 & BNL \\ 
\hline
2	&	4.66 	& 4.88 &	4.11  &   4.11 	&	4.12 	&	4.79 	&	4.16   &   4.16 	&	4.09 	&	4.86 		\\
3	&	3.60 	& 3.69 &	2.96  &   2.97 	&	2.97 	&	3.68 	&	3.00   &   3.01 	&	2.96 	&	3.72 		\\
4	&	2.88 	& 2.90 &	2.20  &   2.22 	&	2.21 	&	2.91 	&	2.25   &   2.26 	&	2.21 	&	2.94 	 	\\
5	&	2.37 	& 2.35 &	1.67  &   1.70 	&	1.69 	&	2.41 	&	1.71   &   1.73 	&	1.69 	&	2.39 	 	\\
\hline
MAE &	 	& 0.09 &	0.64  &   0.63   	&	0.63 	&	0.07 	&	0.60   & 0.59   	&	0.64 	&	0.10 	 	\\
\end{tabular*}
\end{ruledtabular}
\end{table*} 

\newpage
\begin{table*}
\caption{\label{table:2} Same as Table {\ref{table:1}}, but for the $^1L_b$ states of $n$-acenes.} 
\begin{ruledtabular}
\begin{tabular*}{\textwidth}{rrrrrrrrrrrr} 
&  \multicolumn{5}{r}{RT-TDDFT} & \multicolumn{4}{r}{LR-TDDFT} \\ 
\cline{4-7}
\cline{8-11} 
$n$-acene & Experiment & CC2 & LDA & PBE & LB94 & BNL & LDA & PBE & LB94 & BNL \\ 
\hline 
2	&	4.13 	& 4.46 &	          &     	 &	 	&	4.61 	&	4.29   &   4.29 	&	4.21 	&	4.64 		\\
3	&	3.64 	& 3.89 &	3.65   &   3.64 	&	3.68 	&	4.03 	&	3.69   &   3.69 	&	3.61 	&	4.07 		\\
4	&	3.39 	& 3.52 &	3.26   &   3.26 	&	3.28 	&	3.68 	&	3.29   &   3.30 	&	3.23 	&	3.70 	 	\\
5	&	3.12 	& 3.27 &	3.00   &   2.99 	&	3.01 	&	3.42 	&	3.03   &   3.03 	&	2.97 	&	3.44 	 	\\
\hline 
MAE &	 	& 0.22 &	0.09  & 0.09     	&	0.09 	&	0.37 	&	0.10   & 0.10   	&	0.11 	&  0.39 	 	\\
\end{tabular*}
\end{ruledtabular}
\end{table*}

\newpage

\newpage

\newpage

\newpage

\newpage

\end{document}